\documentclass[twocolumn,twoside,slac_two]{revtex4}
\usepackage{graphicx}
\usepackage{fancyhdr}
\begin{document}

\title{Varying Faces of Photospheric Emission in Gamma-ray Bursts}

\author{Magnus Axelsson}
\affiliation{KTH Royal Institute of Technology, Oskar Klein Center, 106 91 Stockholm, Sweden}
\author{\vspace{2mm} on behalf of the {\it Fermi}-LAT Collaboration}
\noaffiliation

\begin{abstract}
Among the more than 1000 gamma-ray bursts observed by the {\it Fermi Gamma-ray Space Telescope}, a large fraction show narrow and hard spectra inconsistent with non-thermal emission, signifying optically thick emission from the photosphere. However, only a few of these bursts have spectra consistent with a pure Planck function. We will discuss the observational features of photospheric emission in these GRBs as well as in the ones showing multi-component spectra. We interpret the observations in light of models of subphotospheric dissipation, geometrical broadening and multi-zone emission, and show what we can learn about the dissipation mechanism and properties of GRB jets.
\end{abstract}

\maketitle

\thispagestyle{fancy}

\section{INTRODUCTION}

Despite having been studied for well over 20 years, the emission mechanisms active during the prompt phase in gamma-ray bursts (GRBs) remain 
unclear. A robust prediction of the fireball model for GRBs\,\cite{cr78,rm94} is that the relativistic jet is initially opaque and therefore photospheric 
emission is inevitable. Yet its strength is uncertain and it is therefore not necessarily detectable. In 1986, both Paczynski\,\cite{pac86} and 
Goodman\,\cite{good86} suggested a strong contribution of photospheric emission in GRB spectra; however, the observed spectra generally appear nonthermal and these models were therefore not considered viable. 

Interest in the photospheric component resumed with observations of GRBs using {\it Compton Gamma-Ray Observatory}/BATSE (20--2000 keV). 
Ryde\,\cite{ryd04} found that in many individual emission pulses an equally good or better fit could be found by using a model comprising a Planck function 
and a power-law, as compared to the traditional Band function. Additionally, it was found that the evolution of the Planck function component during the 
prompt phase followed well defined and consistent characteristics. The Planck component was interpreted as the photosphere of the GRB. At present
there is again mounting evidence from theoretical considerations that the photosphere of the relativistic outflow (jet) plays an important role\,\cite{laz10,vurm11,gia11,zha11}. 

In this paper the observational signs so far attributed to photospheric emission will be discussed and interpreted in light of models of 
subphotospheric dissipation, geometrical broadening and multi-zone emission. Photospheric emission can give rise to many different spectral 
shapes, and pure blackbody emission is rarely expected.

\section{OBSERVATIONS}

As noted above, predictions of photospheric emission came early in the study of GRBs. Yet it was not until the detailed spectral
studies made possible by {\it CGRO}/BATSE that the first clear observational signs were seen. In part this may be due to the 
ambiguity in attributing spectral components to distinct physical processes. This has to some extent meant that the search for
photospheric emission has become a search for blackbody (or Planckian) components in the spectrum: while the photosphere
can in principle give rise to many different shapes, a blackbody can only come from the photosphere.

\subsection{Blackbody-like spectra}

\citet{ghi03} first reported the presence of a blackbody component in the initial phase of some GRBs detected with {\it CGRO}/BATSE.
\citet{ryd04} also showed that some GRBs could be well fit with single Planck functions throughout the prompt phase. However, such
cases are extremely rare. In the entire BATSE catalogue, only 6 out of $\sim2200$ GRBs are well described by a pure blackbody. The situation
is similar for the {\it Fermi} catalogue: only 2 such bursts in over $1400$ have reported \citep{ghi13,lar15}. 

Although these numbers may seem low, what is perhaps more surprising is that there are such cases at all. Already from the start,
it was shown that purely geometrical considerations meant that photospheric emission should be somewhat broader than a single
temperature Planck function. The fact that there are such narrow spectra is thus very constraining for theoretical models.

An interesting case for the study of photospheric emission is GRB090902B, one of the brightest bursts seen by {\it Fermi}. During the first part of the emission
episode, the main spectral peak is very narrow and well-fit by a multicolor blackbody \citep{ryd10}. However, during later times in the 
pulse the spectrum broadens considerably. As the spectral evolution can be followed, it is clear that the same component is seen throughout
the prompt phase. The blackbody-like spectrum at early times ties it to the photosphere, and GRB090902B thus shows that photospheric 
emission is not necessarily Planckian.

\begin{figure}
\includegraphics[width=80mm]{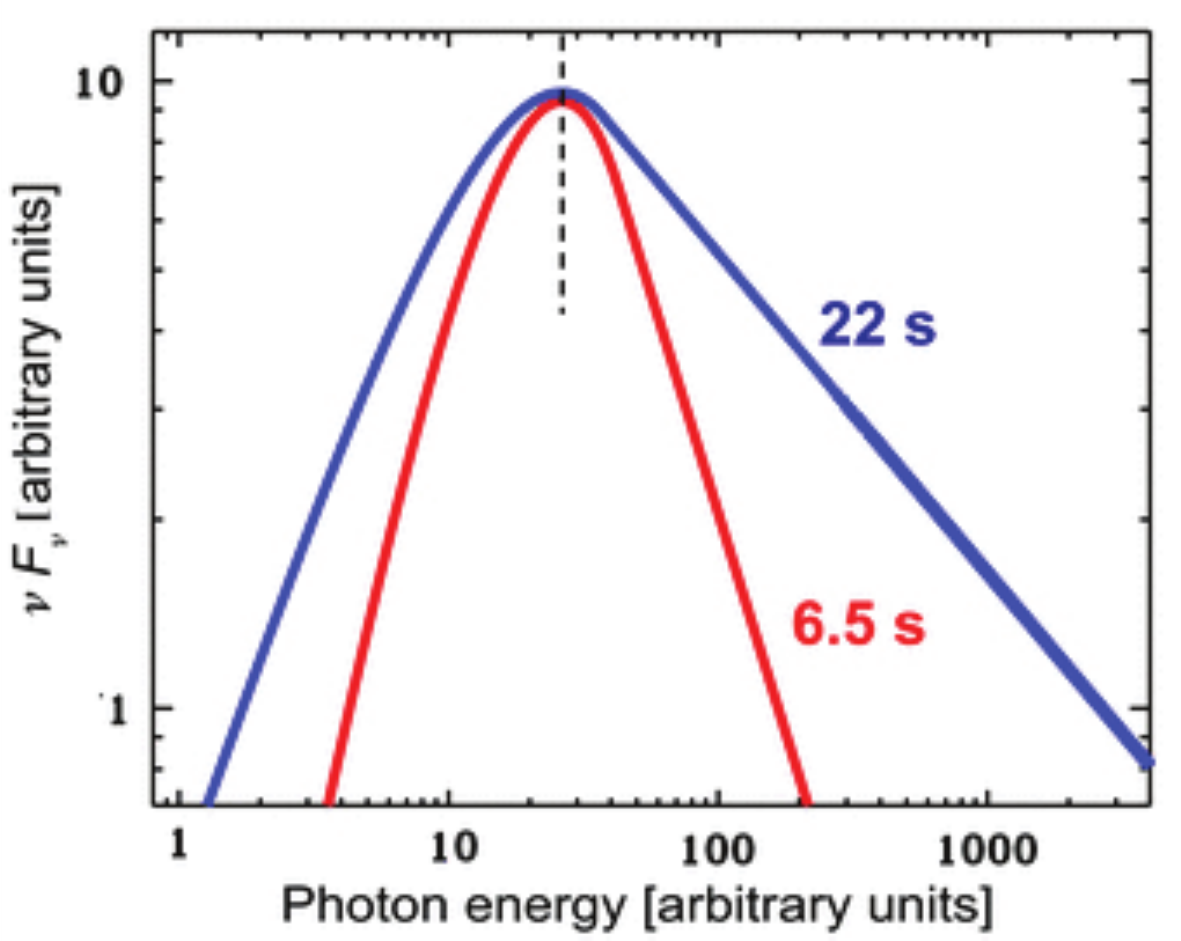}
\caption{Peak-aligned spectra from GRB090902B at two different times, showing the change in width of the spectral peak. At early times in 
the pulse (red), the spectrum is very narrow and well described by a multicolor blackbody. Later in time (blue), the spectrum has significantly
broadened, and is well fit by a Band component with typical parameters.}
\label{090902B}
\end{figure}

In summary:
\begin{itemize}
\item Observations of ``blackbody-like spectra"  indicate that the photosphere is detected, and thereby also plays 
a role in GRB prompt emission and the formation of spectra.
\item The fact that some spectra are well fit by single-temperature blackbodies has strong theoretical implications on
the physical conditions of the emission region. 
\item The slightly wider spectra allow us to probe the broadening mechanisms active in the outflow. This is particularly
true for bursts such as GRB090902B where width changes strongly during the pulse.
\end{itemize}

However, most spectra are not well described by a single narrow component. Nevertheless, evidence of photospheric 
emission in some GRBs motivates us to search for its presence also in other bursts. 


\section{MULTI-COMPONENT SPECTRA}
One of the most striking results of the {\it Fermi} satellite is the discovery of multiple components in the spectrum of GRBs. 
Bright bursts, where the signal-to-noise ratio is highest, show statistically significant deviations from a simple Band function 
\citep{catalog}. One component commonly found is a power-law extending to high energies (e.g., GRB 080916C). However, a
few bursts also show features at lower energies ($\leq 100$ keV), which are well-fit by a Planck function. Perhaps the 
strongest such example is GRB110721A, where the significance of the extra component was greater than $5\sigma$ 
\citep{axe12}.

\begin{figure}
\includegraphics[width=8.5cm]{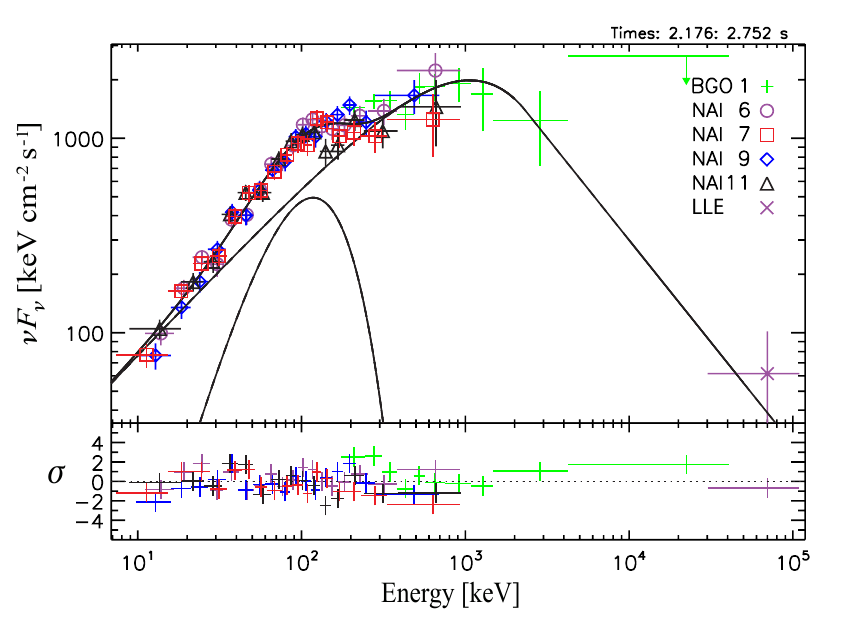}
\caption{Spectrum from GRB110721A, showing the clear detection of an extra component in addition to the Band function.
This component was fit using a blackbody and interpreted as photospheric emission in \citet{axe12}. \label{110721}}
\end{figure}

The results found with {\it Fermi} match those previously seen in BATSE data. \citet{ryd04} found that a model comprising
a blackbody and a power-law provided a good fit to several GRB spectra observed by BATSE. The power-law index was
greater than -2, so it was clear that there had to be a turn-over at higher energies. With the much broader energy range 
afforded by {\it Fermi}, the power-law seen in the BATSE data is revealed as the low-energy slope of the Band component.
It should be noted that also {\it Fermi} has detected a power-law component in the spectra; however, this feature is seen
in addition the the Band component, and the temporal behavior is very different. Its origin is not yet understood, but may
be related to the mechanism producing the temporally extended GeV emission \cite{catalog}.

Another feature which strengthens the common origin of the blackbody components in BATSE and {\it Fermi} spectra
is their temporal evolution. The BATSE components showed a typical behavior where the temperature decayed with time
as a broken power-law. This distinctive feature is also seen in the {\it Fermi} data. 


\subsection{Effects of multiple components}

\begin{figure}
\includegraphics[width=8.cm]{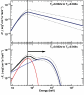}
\caption{Resulting fits when the spectrum from GRB120323 is fit with a pure Band component (top) and an additional
blackbody component (bottom). Adding the blackbody component does not give a statistically sufficient improvement to
claim its presence. However, the parameters of the Band component are changed when the blackbody component is
present, and therefore different interpretations may be allowed. Adapted from \citet{gui13}. \label{peakshift}}
\end{figure}

The additional blackbody component detected is typically subdominant, in general contributing only 5-10\% of the total
flux. For this reason, its presence can only be firmly seen in very bright GRBs. However, it may still be present in many
more GRBs, and this can have important consequences. When spectra are fit with models comprising a blackbody in 
addition to the main Band component, the parameters of the Band component change. This means that although a
photospheric component may not be statistically detected, its presence can have a large impact on the interpretation
of the bulk of the emission. For instance, the peak of the Band component will shift to higher energies, and the measured
value of the low-energy slope, $\alpha$, will soften \cite{gui13}. An example of this is shown in Figure~\ref{peakshift}, where the 
spectrum of GRB120323 is fit with and without an additional blackbody component. The changed parameters may lead 
to the Band component being more compatible with synchrotron emission, and thus change the theoretical
interpretation also of the main emission component.

As thermal emission is a well-known physical process, identifying such a component allows physical parameters of
the outflow to be derived \cite{ryd09}. These include the bulk Lorentz factor, jet-launching radius and saturation radius.
For instance, studies of GRB110721A have found that the Lorentz factor was initially around 1000, then decreased 
throughout the pulse to values $\sim200$ \cite{axe12}. The jet launching radius was instead found to increase from 
$3\times10^6$\,cm to $2\times10^9$\,cm \cite{iyy13}.

\section{INTERPRETATIONS}

In the case of ``typical'' single-component GRB spectra, it is generally assumed that a single process is giving rise to the
emission. For spectra well-fit by a single or multi-temperature blackbody, the most likely candidate is photospheric emission.

For the multi-component GRBs, the interpretation is less straight-forward. A natural first assumption is to connect the two
components to different emission regions. The blackbody component is then attributed to thermal emission arising from the 
photosphere, and the Band component related with non-thermal radiation further out in the jet. There are many different 
possible realizations of the scenario. For instance, the location of the photospheric radius in relation to the saturation radius
will affect the strength of the blackbody and different magnetizations of the outflow will change the ratio between the two
components \cite{dai13}. 

As mentioned above, identifying a blackbody component in the spectrum can alleviate some of the difficulties facing interpretations 
suggesting a synchrotron origin for the Band component. Many observed GRBs have hard spectra below their $\nu F_{\nu}$ peaks.
Those with indices $\alpha > -1.5$ below this peak cannot possess electrons that radiate synchrotron emission in the expected fast 
cooling regime, within this spectral window; this is the so-called fast-cooling $\alpha$ index limit \cite{pre98}. Models 
including a low-energy blackbody component allow for softer slopes of the Band component, thereby making the interpretation more
compatible with synchrotron.

\begin{figure}
\includegraphics[width=8.5cm]{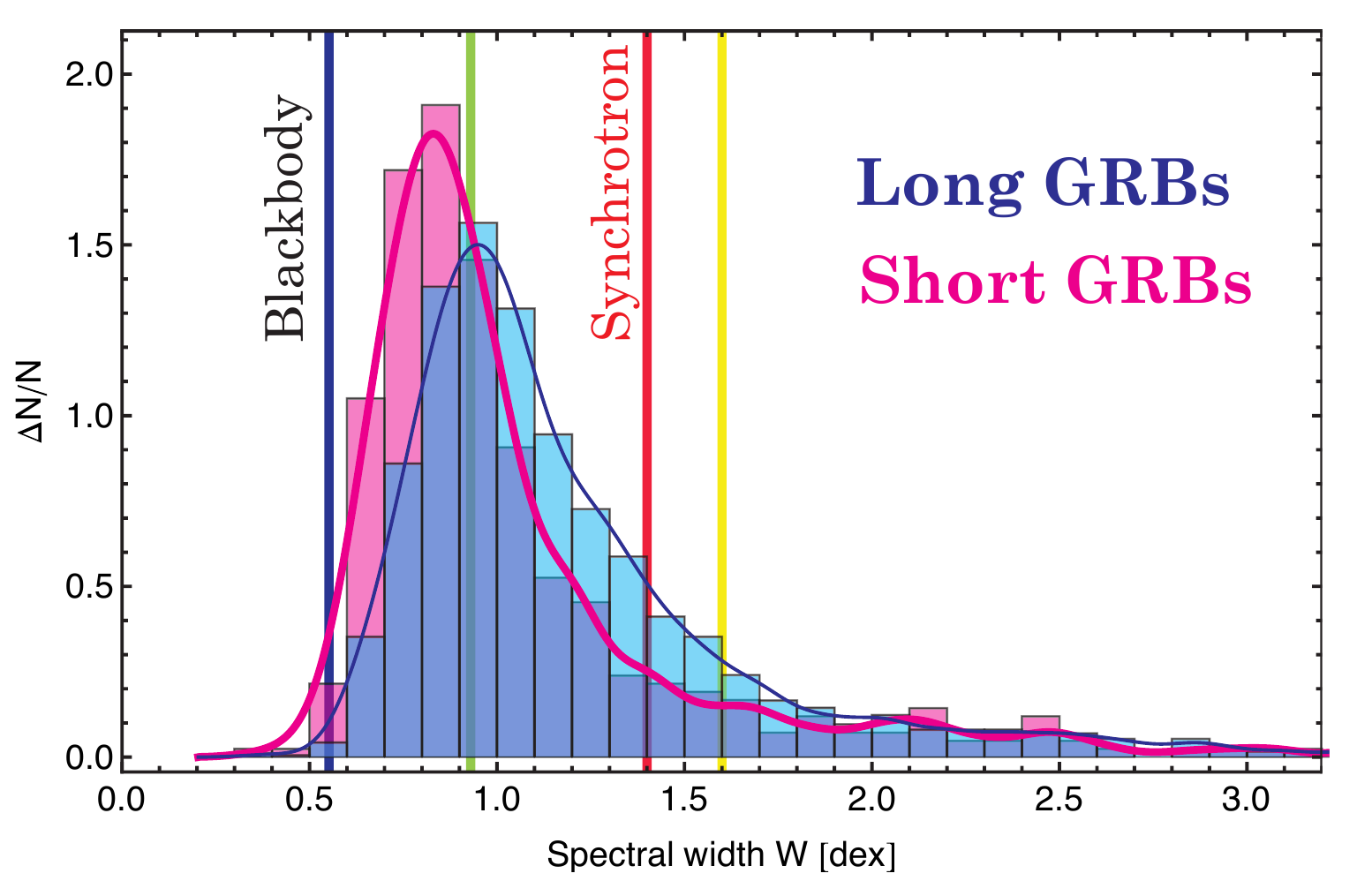}
\caption{Distribution of widths of GRB spectra measured by {\it CGRO}/BATSE and {\it Fermi}/GBM. The solid lines indicate the
width of a blackbody, and synchrotron from three electron distributions: monoenergetic (green), Maxwellian (red) and power-law with 
index -2 (yellow). Adapted from \citet{ab15}. \label{widths}}
\end{figure}

Spectra with hard $\alpha$ slopes are however not the only issue facing synchrotron interpretations. Studying the widths of spectra, 
it can be seen that most are too narrow to accommodate synchrotron emission from realistic electron distributions \cite{ab15}. This
is shown in Fig.~\ref{widths}. In these cases adding a blackbody component will not help, but rather worsen the issue; the width of
the Band function component in a composite spectrum is if anything more narrow than the entire spectrum.

An alternative to multiple emission zones is that the entire spectrum arises from the photosphere. This of course requires a radical
departure from the framework where photospheric emission is described by a (single or multicolor) blackbody. One suggested way of
altering the spectrum is subphotospheric emission. In brief, in this scenario energy is dissipated below the photosphere, modifying the 
emergent spectrum. Different models propose different origins, such as magnetic reconnection \citep{gia08}, internal shocks \cite{ioka10} 
or collisional dissipation \cite{bel10}. By varying the amount of dissipation and parameters of the outflow, it is possible to produce a 
wide range of spectral shapes by such subphotospheric energy release \citep{peer06,nym11}.

As described by \citet{pac86}, geometrical effects will broaden the spectrum. Considering relativistic limb darkening, \citet{lun12} used 
a combination of analytical model and Monte Carlo simulation to study the emergent spectrum from a jet. It was found that for a narrow 
jet, with opening angle is of the order of the relativistic beaming angle, a broadening of the photospheric spectrum is expected for any 
viewing angle. For a broader jet, the broadening effect is strong only if the viewing angle lies along the edge of the outflow, i.e., is close 
to the jet angle. 

Apart from increasing signal to noise in spectra, is there any way to distinguish between these scenarios? Recently, polarimetry
has become the focus of attention, and does provide valuable extra information. In the case of non-thermal emission, synchrotron
radiation is expected to have high degrees of polarization. Yet also photospheric emission can be polarized, though polarization
degrees $\le 40$\% are expected \cite{lun14}. Predictions of correlations between spectra and polarization may thus allow us to
determine the physics behind the prompt phase emission. Unfortunately, there are at present very few reliable measurements of
polarization in GRBs.

\section{CONCLUSIONS}

Photospheric emission has been detected in a growing number of GRBs, with Planckian components appearing together with 
other components, or dominating the spectrum. This shows that the photosphere plays a part in the formation of the spectra.

Most GRB spectra do not look thermal, and many instead having multiple components. This can be interpreted as radiation from
two separate emission regions, or as pure photospheric emission. Understanding the role of the photosphere is thus important
to probe the physics of the outflow itself.

Polarimetry provides a possible way to determine the contribution of the photosphere. There are today several proposed missions
capable of measuring polarization in GRBs, which promises new insight into the physics of the relativistic jet.

\begin{acknowledgments}
The \textit{Fermi}-LAT Collaboration acknowledges support for LAT development, operation and data analysis from NASA and DOE (United States); CEA/Irfu and IN2P3/CNRS (France); ASI and INFN (Italy); MEXT, KEK, and JAXA (Japan); and the K.A.~Wallenberg Foundation, the Swedish Research Council and the Swedish National Space Board (Sweden). Science analysis support in the operations phase from INAF (Italy) and CNES (France) is also gratefully acknowledged.
\end{acknowledgments}

\end{document}